\documentclass{egpubl}
\usepackage{egsgp2020,amsmath,amssymb}
 
\SpecialIssuePaper         

\usepackage[T1]{fontenc}
\usepackage{dfadobe}  

\biberVersion
\BibtexOrBiblatex
\usepackage[backend=bibtex,bibstyle=EG,citestyle=alphabetic,backref=true]{biblatex}
\electronicVersion
\PrintedOrElectronic

\ifpdf \usepackage[pdftex]{graphicx} \pdfcompresslevel=9
\else \usepackage[dvips]{graphicx} \fi

\usepackage{egweblnk} 

\newtheorem{theorem}{Theorem}[section]

\newtheorem{definition}[theorem]{Definition}

\title[Topology-Aware Surface Reconstruction for Point Clouds]{Topology-Aware Surface Reconstruction for Point Clouds}

\author[R. Br\"uel-Gabrielsson, V. Ganapathi-Subramanian, P. Skraba, L. Guibas]{Rickard Br\"uel-Gabrielsson\textsuperscript{1,2} \ Vignesh Ganapathi-Subramanian\textsuperscript{1} \ Primoz Skraba\textsuperscript{3,4} \ Leonidas J. Guibas\textsuperscript{1} \\
\textsuperscript{1}Stanford University \ \textsuperscript{2}Unbox AI \ \textsuperscript{3}Queen Mary University of London \ \textsuperscript{4}Jo\v{z}ef Stefan Institute \\ \texttt{ rbg@cs.stanford.edu, vigansub@stanford.edu, p.skraba@qmul.ac.uk, guibas@cs.stanford.edu}}

\begin{document}


\teaser{
\centering
  \includegraphics[width=0.8\linewidth]{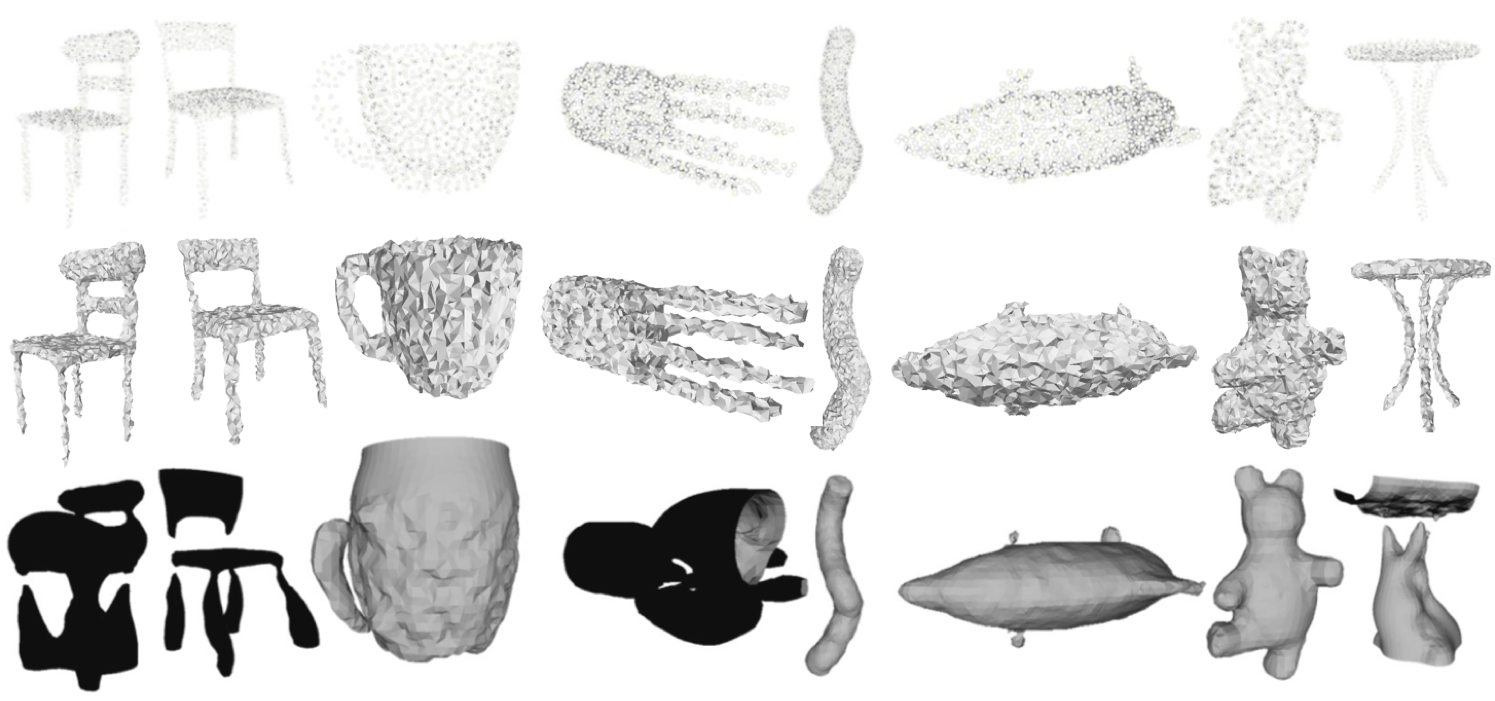}
  \caption{Reconstruction of example point clouds from the McGill dataset~\cite{siddiqi2008retrieving}. \textit{First row:} Input Point cloud. \textit{Second row:} Our reconstruction of the shape in the first row. \textit{Third row:} Poisson surface reconstruction~\cite{kazhdan1poisson} of the shape in the first row. Shapes showcased (left to right) are two chairs, cup, octopus, snake, dolphin, teddy and table respectively.}
  \label{fig:comparison_poisson}
}

\maketitle
\begin{abstract}
  We present an approach to incorporate topological priors in the reconstruction of a surface from a point scan. We base the reconstruction on  basis functions which are optimized to provide a good fit to the point scan while  satisfying predefined topological constraints.  We optimize the parameters of a model to obtain a likelihood  function  over  the  reconstruction domain. The  topological constraints are captured by persistence diagrams which are incorporated within the optimization algorithm to promote the correct topology. The result is a novel topology-aware technique which can (i) weed out topological noise from point scans, and (ii) capture certain nuanced properties of the underlying shape which could otherwise be lost while performing surface reconstruction. We show results reconstructing shapes with multiple potential topologies, compare to other classical surface construction techniques, and show the completion of  real scan data. 


\end{abstract}  
\section{Introduction}
\label{sec:intro}

Shapes are characterized by a number of markers that are representational and explain their different properties. These could be geometric, structural, topological, functional, or stylistic. These properties add to our understanding of shape collections and can be instrumental in solving important problems in 3D vision, geometry processing, and computer graphics. Applications include shape alignment, shape correspondences, surface reconstruction, and shape synthesis.
Of these different considerations, the topological invariance of shapes is an oft-observed trait across collections that is seldom explicitly exploited. It is an important property of many collections, especially those where the shapes have common structure. This phenomenon can be observed in human body parts. For example, the topology of a healthy human heart is always the same, whether it comes from a child or an adult. 
One of the reasons this has not been explicitly addressed is that most shape processing problems, in particular surface reconstruction techniques, involve choices that are both continuous and combinatorial. Continuous choices involve parameter regression, where a certain set of parameters can be regressed from within a continuous set. This could include learning chair leg lengths or the radius of a spherical object. The more computationally difficult choices are combinatorial or discrete  and often involve making decisions about shape  classifications; topological information is of this type.  For example, this appears in regression of shape grammars, where an example of a combinatorial choice is that of having an armrest or not. Combinatorial choices are extremely challenging to make and prove to be a bottleneck in many shape processing tasks.
 
To make topological information more amenable to optimization, we use \emph{persistent homology}. This is a tool that provides topological markers, called persistence diagrams, that capture these combinatorial choices through a continuous proxy,  describing the topology of point clouds over multiple scales. Surface reconstruction from a point scan usually involves being oblivious to the scale of the point scan which can lead to topologically incorrect reconstructions. We present a technique to extract a surface from a point scan while preserving the predefined topology of the shape. Since persistence-based tools range across different scales, we manage to filter out a reliable reconstruction of the point scan which respects the requisite topology. 

Most surface reconstruction techniques manage topology in a post-processing step, such as removing spurious components  in the reconstruction or ensuring that an extracted surface is \emph{watertight}. 
However, topological information can aid in completing sparse point scans reliably, as well as resolve topological ambiguities, especially in the case of non-intersecting close regions on the surface of the shape. We highlight the advantages of topology-aware shape completion with multiple examples in Section~\ref{sec:results}.
Our technique constructs likelihood functions that take high values on points that likely lie on or near the surface. These functions are informed by backpropagation from the persistence diagrams of candidate surfaces, aiding to obtain one of requisite topology. These diagrams act as an intermediary between continuous scale information around the points of the scan and the combinatorial nature of topology. 
A summary of our contributions is:
\begin{itemize}
    \item A novel topology-aware likelihood function, optimized on a point cloud, and based on persistent homology measures of any dimension.
    \item An  automated surface reconstruction algorithm that may preserve or even create the requisite topology.
    \item A new measure of topological fidelity for reconstruction along with a comparison of our approach with existing methods.
\end{itemize}

\section{Related Work}
\label{sec:related}

Reconstructing surfaces from point clouds is a difficult problem which has received extensive attention.  We first describe related work in surface reconstruction, followed by an overview of related topological techniques.

\subsection{Surface Reconstruction from Point Clouds}
\label{subsec:surfacerecons}
The problem of surface reconstruction has multiple facets and many interesting sub-problems, we direct the reader to the survey by Berger et. al.~\cite{berger2014state} for further reading. 
Surface reconstruction from point clouds appears in numerous scenarios including urban reconstruction~\cite{musialski2013survey} and  completing partial surfaces or point scans~\cite{attene2013polygon}. The specfic problem we consider is interpolating a point cloud to generate a surface~\cite{hoppe1992surface,guennebaud2007algebraic,dey2006curve}. 

Most surface reconstruction techniques model the surface to be characterized by the zero-level set of a function that is defined over space~\cite{kazhdan1poisson,carr2001reconstruction,samozino2006reconstruction}. These works approximate the signed Euclidean distance from the underlying surface, and essentially interpolate values between the input points, extracting the underlying surface as a zero set of this interpolated distance field. Since there is no additional information in this sort of technique, there is nothing to prevent points far from the surface to have extremely low values. Therefore, many techniques use an \textit{inside-outside points} approach where points inside the surface and outside it are provided so as to guide the function sign at various regions in space. Moreover, while this is an extremely elegant representation of a shape, it is also restrictive, with the optimization step including multiple equality constraints. The work by Poranne at al.~\cite{poranne20103d} takes a minor detour from this, where they do not necessarily force the surface value to be zero. Instead, they obtain the surface by applying an algorithm based on the watershed transform~\cite{roerdink2000watershed} which extracts the low-level set of the function values as the surface.  In this work, we build a likelihood function where points on the surface are given high values, with no value constraints. This way, the surface values can develop organically. A simple surface reconstruction technique from existing function values can then be used to obtain the actual surface.

Topological control in surface reconstruction has been performed by user interaction~\cite{sharf2007interactive,yin2014morfit}, fixed template optimization~\cite{bazin2005topology,gs2018parsing,zeng2008topology}, removing topological errors from an existing surface~\cite{ju2007editing,wood2004removing}, and optimization-based  surface reconstruction~\cite{sharf2006competing,huang2017topology,zhou2014topology,lazar2018robust}.  In this work, we approach the topology-guided reconstruction along the lines of the optimization-based technique, locating parameters that maximize the function values at surface points, while preserving the topology, using ideas from persistent homology as discussed below. 

\subsection{Persistent Homology}
\label{subsec:homology}
Persistent homology, or simply persistence, is a well-established tool in applied and computational topology.
%
Topological simplification based on function optimization is present in some of the earliest work on persistence~\cite{edelsbrunner2000topological}. More generally, it been used as a tool to satisfy topological criteria in a variety of geometry processing applications including shape matching ~\cite{carlsson2005persistence}, optimal pose-matching~\cite{dey2010persistent}, and shape segmentation~\cite{skraba2010persistence}.
The problem of topological simplification of shapes  while preserving persistent features in given data has traditionally been performed in the context of function denoising~\cite{attali2009persistence,bauer2012optimal}. The inclusion of topological information in optimization has appeared in the work of Gameiro et al.~\cite{gameiro2016continuation} where the authors attempt to perform point cloud continuation for dynamical systems based applications. Recent work by Poulenard et al.~\cite{poulenard2018topological} use the idea of function optimization for the purpose of shape matching. The work in this paper mirrors that of Poulenard et al. by performing function optimization on basis coefficients, which are used as the building blocks of a given function, which in our case is representative of the shape surface.  This general approach has received a large amount of attention recently including theoretical analysis~\cite{leygonie2019framework} as well as numerous applications~\cite{bruel2019topology}. We refer the reader to ~\cite{leygonie2019framework} for a recent overview of applications.

\noindent{\textbf{Overview: }}
The rest of the paper is arranged as follows: we begin with a discussion of classical surface reconstruction in Section~\ref{sec:surface}, followed by the  required topological preliminaries in Section~\ref{sec:persistence}.  The two ideas are combined to perform topology-aware surface reconstruction. In Section~\ref{sec:top_opt}, we discuss the optimization to find locally optimal parameters to a topology-aware likelihood function. Once the likelihood function has been computed, we extract a surface using the technique described in Section~\ref{sec:surface_recon}. Section~\ref{sec:results} showcases the technique in practical example cases and evaluates it against other state-of-the-art reconstruction techniques. Section~\ref{sec:conclusion} concludes the paper with a brief discussion on future work.

\section{Implicit Surface Representation}\label{sec:surface}
Using level sets of functions to represent surfaces  reconstructed from point scans is a well-established technique ~\cite{kazhdan1poisson,carr2001reconstruction}. At a high level, previous techniques optimize the function so that the level set and hence the extracted surface closely approximate the point cloud. We perform the optimization while promoting the correct topology of the extracted surface.  

Classical surface reconstruction techniques use different approaches to build a function to fit a surface to a point scan. 
Poisson reconstruction constructs a function which evaluates to $0$ on the surface. It requires normal information to label points as inside and outside the surface, assigning  $+1$ to points outside the surface and $-1$ inside the surface. This can be used to then reduce the problem to solving the Poisson equation~\cite{kazhdan1poisson}. Alternatively, radial basis functions can be used, again evaluating to $0$ on the surface and a small value at nearby points not on the surface~\cite{carr2001reconstruction}. 
To avoid confusion, we note that in our approach, we attempt to build the surface to be higher-valued as opposed to lower-valued as in the above two techniques. 
We construct multivariate Gaussians centered around the points in the scan, and then minimize a topology-aware loss function using stochastic gradient descent to optimize the covariance matrices of the Gaussians. We then use a topology-aware surface reconstruction technique to obtain the surface. 
More generally, we consider basis functions $\varphi_p$ that are defined for every point in the point cloud,   $p\in\mathcal{P}$. Each function is parameterized by a vector $\alpha_p$. The likelihood function of the surface including a point $x \in \mathbb{R}^{d}$ is a linear combination of these basis functions:
\begin{equation}
f(x, \alpha_\mathcal{P}) = \sum_{p \in \mathcal{P}} \varphi_p(x, \alpha_p)
\label{eq:basis_comb}
\end{equation}
where $\alpha_\mathcal{P}$ refers to the collection of parameters $\alpha_p$ for all points $p \in \mathcal{P}$.
The choice of basis functions $\varphi_p$  are a design choice, and are ideally functions whose maxima are at the point $p$ around which they are centered.

\begin{figure}[t]
\includegraphics[width=\columnwidth]{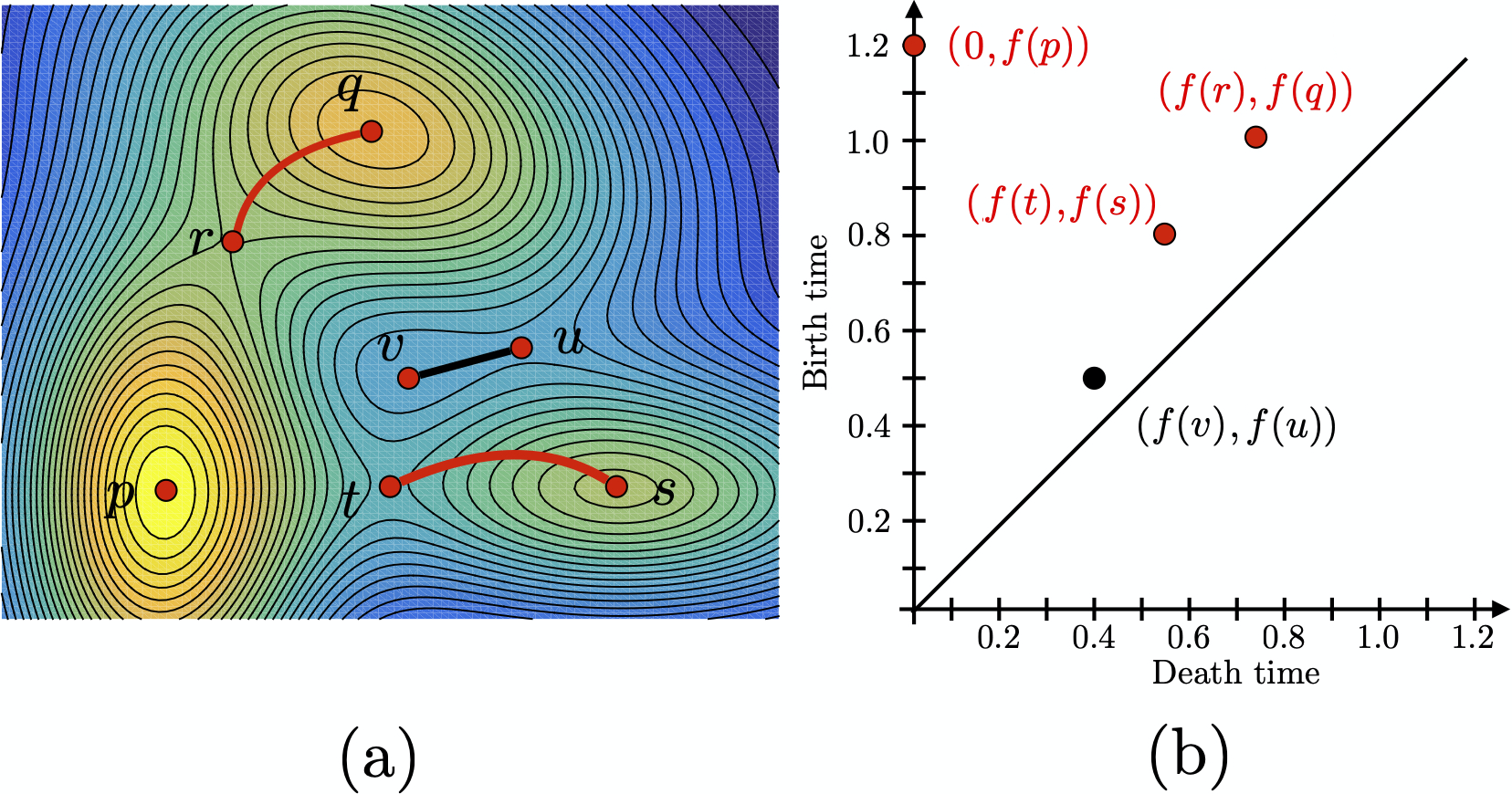}
\caption{\label{fig:persistence} An example of a persistence diagram for the sum of three Gaussians in 2D. (a) Heat map of the function, and (b) The corresponding persistence diagram. The critical points of the height function $p$, $q$, and $s$ are local maxima, $r$, $t$, and $u$ are saddles, and $v$ is a local minimum. Each of the local maxima creates a component (the red points on the right represent 0D homology classes where the heights are the birth times). Saddles $r$ and $t$ merge components and correspond to the death times of two of the red points. The saddle $u$ creates a ring (1D homology class), which is closed when the minimum at $v$ is reached, corresponding to the black point in (b). }
\end{figure}
 As mentioned above, we use multivariate Gaussians as our basis functions. A Gaussian centered at a point $\mu$ is given by 
$$G(x;\mu,\Sigma) = \sqrt{{(2\pi)}^{-3}|\Sigma|^{-1}}\textbf{exp}(-(x-\mu)^T\Sigma^{-1}(x-\mu))$$
where $\Sigma$ is a symmetric covariance matrix. We define the Gaussian basis-function around point $p$ of $\mathcal{P}$ to be $\varphi_p(x;\alpha_p) = G(x;p,\alpha_p^T\alpha_p)$. 
The covariance matrix is a symmetric positive definite matrix
and hence can be decomposed as $M = A^TA$ for some matrix $A$. The singular value decomposition (SVD) of $M$ is given by $M = U\Lambda U^T$, defining $A = U\Lambda^{1/2}U^T$. If $\Lambda = \textbf{diag}(\lambda_1, \lambda_2, \lambda_3)$, then $\Lambda^{1/2} = \textbf{diag}(\sqrt{\lambda_1}, \sqrt{\lambda_2}, \sqrt{\lambda_3})$. We define our parameters to be the symmetric square roots of the covariance matrix, i.e. $\alpha_p=A$.  

The parameters of the likelihood functions, that is the covariance matrices, are initialized such that the derivative of the Gaussian is maximized at the average distance between points. 
To improve computational efficiency we only evaluate the likelihood function on a grid, which we denote by $\mathcal{X}$. 

In the rest of the paper, we use the notation $p\in \mathcal{P}$ to denote points in the point cloud and $x\in \mathcal{X}$ to denote the grid points where the function is evaluated. The resolution of the grid ensures non-trivial distances between points in the point cloud and the grid points.  Rather than use a uniform grid, we enlarge $\mathcal{X}$ by treating the likelihood function as a probability distribution over a much finer grid from which we sample additional points. This allows us to have a higher grid-resolution around the points $\mathcal{P}$ and so capture more detail without significantly increasing computational costs. 



In the following sections, we build on the framework from ~\cite{poulenard2018topological} to derive a backpropogation function to optimize the likelihood function with respect to topological constraints.


%

\section{Topological Preliminaries}
\label{sec:persistence}
Our main tool to promote certain topological structures is \emph{persistent homology}. Here we review the relevant topological notions which will be used in later sections. For readers who are unfamiliar with persistence we provide some intuition; however, for a more complete introduction to the underlying theory,  we refer the reader to 
\cite{books/daglib/0025666}. We focus on our specific setting, although many of the techniques readily generalize.  

The idea of the likelihood function, $f(\cdot,\alpha_\mathcal{P})$, is that it should evaluate to larger values close to surface. The super-level set $f^{-1}([t,\infty), \alpha_\mathcal{P})$ should therefore contain the surface with the correct topology for some known $t$. Hence our goal is to optimize $f$ so that super-level sets of $f$ both fit the points and have the correct topology.   

To study the topology of a space, we first construct a combinatorial representation of that space. This is done by building a triangulation of the space in the form of an embedded \emph{simplicial complex} $K$ where the vertices are the points $\mathcal{X}$. Recall that a $k$-dimensional simplex is the convex combination of $k+1$ vertices (or equivalently, $k+1$ points in $\mathcal{X}$). We restrict ourselves to $\mathbb{R}^2$ and $\mathbb{R}^3$, and so we only need to consider vertices, edges, triangles, and tetrahedra. 
As previously mentioned, for computational reasons we only compute $f$ on a set of points $x\in \mathcal{X}$. We initialize the vertices of $K$ as the points of $\mathcal{X}$ and construct a Delaunay triangulation to obtain the higher dimensional simplices. 
The function $f$ on $\mathcal{X}$ can be extended to $K$ in a piecewise-linear fashion. In the following, we do not distinguish between the PL-approximation and the true function, but we address this at the end of this section. 

\begin{figure}
     \centering
     \includegraphics[width=\columnwidth]{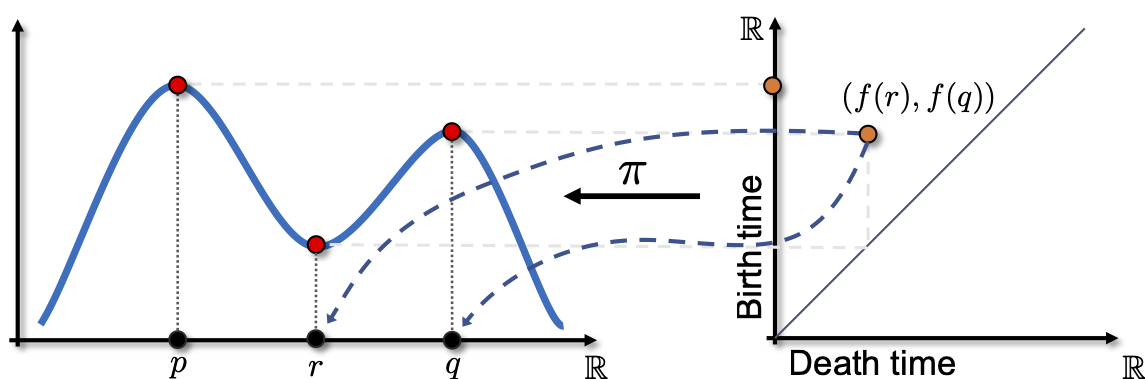}
     \caption{A one dimensional example of a persistence diagram and the inverse map $\pi$. The function on the left has critical points at points $p$, $r$, and $q$. The local maxima create components in the super-level sets and so represent birth times ($y$-axis), while the minimum kills one of the components (the younger one) and so is a death time ($x$-axis). The inverse map for a point in the diagram returns the corresponding critical points/simplices. }
     \label{fig:overview}
 \end{figure}

To describe the topology, we use \emph{homology}, which captures certain aspects of connectedness of a space. For completeness, we define homology as we refer to it at the end of this section. For a simplicial complex $K$, we can consider the vector spaces generated by the $k$-simplices, with one vector space per dimension, denoted by $C_k(K)$, i.e. the chain groups constructed with field coefficients. One can define a boundary operator, which is a linear map $\partial_k: C_k(K)\rightarrow C_{k-1}(K)$ such that $\partial_{k}\circ \partial_{k+1} = 0$. The $k$-dimensional homology is defined as 
$$
 H_k (K) = \frac{\mathrm{ker}\; \partial_{k} }{\mathrm{im}\; \partial_{k+1}} $$
where elements of $\mathrm{ker}\; \partial_{k}$ are called $k$-cycles and elements of $\mathrm{im} \;\partial_{k+1}$ are called $k$-boundaries. 
%
Two cycles are \textit{homologous} if their difference can be written as a linear combination of boundaries, and the set of all cycles that are homologous to a given cycle is called a \textit{homology class}. 
 The rank of the $k$-th homology counts the number of $k$-dimensional features, i.e. 0-dimensional features are connected components, 1-dimensional features are holes, and 2-dimensional features are voids.  We cannot optimize directly for the correct homology as the rank or number of holes is a discrete quantity. Rather, we use \emph{persistent homology} which tracks how homological features appear and disappear over a filtration, i.e. a sequence of spaces related by inclusion. In our setting, we consider the filtration induced by the super-level sets of the likelihood function $f: K \to \mathbb{R}$. Defining,
$$K^\alpha  = \{\sigma\in K \ | \ \forall v \in \sigma, f(v)\geq \alpha \}$$ 
This defines a filtration since $K^a\subseteq K^b$ for all $a\geq b$. Surprisingly, the homology of filtrations can be fully described by the appearance and disappearance of features, called births and deaths. The set of the pairs of births and deaths are called a \emph{persistence diagram}. We omit the formal algebraic definition, but for our purposes we can define a persistence diagram as a map from a space and a function to a set of points in $\mathbb{R}^2$.
$$\mathrm{PD}_f(k): (K,f) \rightarrow \{b_i,d_i\}_{i\in\mathcal{I}_k}$$
where $\mathrm{PD}_f(k)$ refers to the births and deaths of $k$-dimensional homological features, and we drop subscript $f$ when the function $f$ is clear from the context. We define $\mathrm{PD}_f = \cup_k \mathrm{PD}_f(k)$. We often refer to the collection of points as the persistence diagram, with the implication that $K$ and $f$ are fixed; see Figures \ref{fig:persistence} and ~\ref{fig:overview} for examples in 1D and 2D (where $\mathrm{PD}(0)$ and $\mathrm{PD}(1)$ are superimposed) respectively.

 Crucially, it is possible to define meaningful distances between diagrams. This makes it possible to talk about \emph{how far} a given space is from the desired topology, which in turn allows for the  optimization of topology over persistence diagrams. There is a large literature on distances between diagrams and their respective properties \cite{cohen2007stability,cohen2010lipschitz, bubenik2015statistical}. However, we do not use these distances directly; instead we introduce specific cost functions used in Sections~\ref{sec:top_opt} and \ref{sec:results}. 
 
A key ingredient is the existence of an inverse map from the points of the persistence diagram back to $K$
$$ \pi'_f :   \{b_i,d_i\}_{i\in\mathcal{I}_k} \rightarrow (\sigma,\tau)$$ 
 where $\sigma,\tau \in K$. This map is formally defined and  used in ~\cite{poulenard2018topological,leygonie2019framework}. Intuitively, this map can be understood algorithmically. For each homological feature, there is one simplex which creates it and one simplex which bounds it (or kills it); $ \pi'_f $ is simply this correspondence. 
%
In the setting of a super-level set filtration, the function value of a simplex is given by a vertex value:
$$f(\sigma) = \min\limits_{v\in \sigma} f(v)$$
Hence, we can refine the inverse map to
$$ \pi_f :   \{b_i,d_i\}_{i\in\mathcal{I}_k} \rightarrow (x_{b,i},x_{d,i})$$ 
 where $x_{b,i},x_{d,i}\in \mathcal{X}$ are  the extremal vertices in corresponding simplices, see Figure~\ref{fig:overview}.
 
 The final concept we define is the \emph{cycle representative} of a homology class which is our initial candidate for the reconstructed surface.  For simplicity, we assume homology with $\mathbb{Z}_2$ coefficients since we use the support of the cycle, i.e. all simplices with non-zero coefficients.  A \emph{cycle representative} of a homology class is a cycle which represents the entire homology class. This is precisely equivalent to choosing a basis element for a vector space. 
 
In computing persistent homology, most algorithms compute a cycle basis\footnote{Some algorithms only output the barcode or use cohomology, however the standard algorithm \cite{books/daglib/0025666} returns a cycle bases.}. This cycle is \emph{canonical} under some genericity assumptions as it equivalent to a minimum weight basis   \cite{skraba2017randomly}. In practice, the cycle is stable to small perturbations although it is not difficult to construct cases where it is not. 
We refer this cycle as the \textit{Persistent Cycle Representative} (PCR) of a point in the persistence diagram. This cycle most often serves as our initial estimate of the surface (see Section~\ref{sec:surface_recon}). This works well when the desired surface is a manifold and so there is a unique top-dimensional homology class which captures the surface. When there are multiple top-dimensional classes, as in the wedge of two circles as in Figure~\ref{fig:top_guarantee_ring}, we return the union of the PCRs corresponding to the top-dimensional classes. In cases where there is no top dimensional class as in Figure~\ref{fig:heart}, we output the super-level set itself as a thickened surface -- see Section~\ref{sec:surface_recon} for a further details.

We conclude this section with the following remarks. For computational efficiency, we use a PL-approximation of the likelihood rather than the true function. For this approximation, there is an equivalence between super-level set and upper-star filtrations which we use implicitly when defining the inverse map. Furthermore, using the stability of persistence diagrams, it is possible to rigorously bound the error in the persistence diagrams introduced by this approximation in terms of the chosen grid spacing. We omit this as the derivation and proof would introduce additional technicalities without adding much substance to our results, as there are no stability results for the inverse maps and cycle representatives.
We note the work of \cite{edelsbrunnerfasy2013}, which showed that there do exist specific configurations where the error could be non-negligible, but that these do not occur in practice.
%
Finally, we do not discuss the algorithms for computing persistence diagrams as it has been extensively studied, with a number of efficient implementations available. We direct the reader to~\cite{otter2017roadmap} for a recent survey.

\section{Topology-aware Optimization}
\label{sec:top_opt}

Our optimization procedure assumes that we have prior information about the topology of our surface. Determining the topology from a point cloud is a related but separate problem, which we do not address here.  
 We build on the framework from ~\cite{poulenard2018topological} to optimize the parameters of the likelihood function  $\alpha_\mathcal{P}$. Specifically, we derive a backpropogation function to compute the gradient with respect to a topological prior. 
In this section, we consider the parameterized function  $f(\alpha_{\mathcal{P}}): K \rightarrow \mathbb{R}$ with the associated persistence diagram $\mathrm{PD}_{f(\alpha_{\mathcal{P}})}$ and inverse mapping $\pi_{f(\alpha_{\mathcal{P}})}: (b_i,d_i) \rightarrow  (x_{b,i},x_{d.i})$ where $x\in \mathcal{X}$. 

We define a topological prior as a functional on the space of diagrams:
\begin{equation}
    \label{eq:functional}
\mathcal{E}(\mathrm{PD}_f): \{(b_i,d_i)\}_{\mathcal{I}} \rightarrow \mathbb{R}  \end{equation}
In general, the functional can take in persistence diagrams for all dimensions but we often restrict to a single fixed dimension. 

From this point on, we assume that the index $\mathcal{I}_k = \{1,\ldots, N_k\}$ is sorted by decreasing persistence. In other words, $|d_i-b_i| > |d_j-b_j|$ for $i<j$. We also note that $\mathcal{I}_k$ is not static throughout the optimization process but recomputed for every new changed function $f$. 
For a reconstruction, a topological prior is expressed via Betti numbers. That is, how many components we would like the reconstruction to have, how many holes, or voids. In principle, we could include higher dimensional information, but here we focus on surfaces in $\mathbb{R}^3$. Hence, if we would like the reconstruction to have $\ell$ $k$-dimensional features, the functional we use emphasizes the $\ell$-most persistent $k$-dimensional features (and de-emphasizes the less peristent features, i.e. tries to send them to the diagonal) in the reconstruction is:
 \begin{equation}\label{eq:example_func}
 \mathcal{E}(\mathrm{PD}_{f(\alpha_\mathcal{P})}(k)) = -((d_{\ell}-b_{\ell})^2- (d_{\ell+1}-b_{\ell+1})^2),
 \end{equation}
Or in other words, it maximizes the difference between the lifetime of the $\ell$-th most persistent feature and the $(\ell+1)$-th most persistent feature. This does not guarantee convergence to a "reasonable" solution, i.e. we have no guarantees about the resulting features, but in practice it converges to what one would expect.
 
There are many factors to consider when creating topological priors and we describe precise functionals in Section~\ref{sec:results}. Here we derive the general formula, assuming the functional is of the form given in Equation~\ref{eq:functional}. Using the chain rule, 
\begin{align}
    \frac{\partial\mathcal{E}}{\partial\alpha_\mathcal{P}} &= \sum\limits_{i\in \mathcal{I}} \frac{\partial\mathcal{E}}{\partial b_i} \frac{\partial{b_i}}{\partial \alpha_\mathcal{P}}  + \frac{\partial\mathcal{E}}{\partial d_i} \frac{\partial d_i}{\partial \alpha_\mathcal{P}} \nonumber\\
    &=\sum\limits_{i\in \mathcal{I}} \frac{\partial\mathcal{E}}{\partial b_i} \frac{\partial{f(x_{b,i})}}{\partial \alpha_\mathcal{P}}  + \frac{\partial\mathcal{E}}{\partial d_i} \frac{\partial f(x_{d,i})}{\partial \alpha_\mathcal{P}} \label{eq:general}
\end{align}
where we use $\pi_{f(\alpha_\mathcal{P})}(b_i,d_i) = (x_{b,i},x_{d,i})$. In our setting,
$$f(x,\alpha_\mathcal{P}) = \sum_{p\in\mathcal{P}}G(x;p,\alpha^T_p\alpha_p),$$ 
and hence 
\begin{multline*}
\frac{\partial f(x,\alpha_\mathcal{P})}{\partial \alpha_\mathcal{P}} = \sum_{p \in \mathcal{P}} f(x, \alpha_p) \left( -\frac{1}{2}\left((\alpha_p^T\alpha_p)^{-1}\right.\right.\\
\left.\left. -(\alpha_p^T\alpha_p)^{-1}(x-p)(x-p)^T(\alpha_p^T\alpha_p)^{-1}\right)\frac{\partial \alpha_p^T\alpha_p}{\partial \alpha_p} \right)
\end{multline*}
where
$$ \frac{\partial \alpha_p^T\alpha_p}{\partial \alpha_p} = \alpha_p^T J^{ij} + J^{ji}\alpha_p \text { such that } (J^{ij})_{kl} = \delta_{ik} \delta_{jl} $$
This can be evaluated at $x_{b,i}$ and $x_{d,i}$ and substituted into Equation~\ref{eq:general}. It remains to compute the derivative of the functional to complete the formula. For example given the functional in Equation~\ref{eq:example_func}, we obtain 
\begin{multline*}
\frac{\partial\mathcal{E}}{\partial \alpha_{\mathcal{P}}} = \sum_{p \in \mathcal{P}} (\alpha_p^T\alpha_p)^{-1} \big( D(\alpha_p, p,x_{b,\ell},x_{d,\ell},b_{\ell},d_{\ell})\\-D(\alpha_p, p,x_{b,\ell+1},x_{d,\ell+1},b_{\ell+1},d_{\ell+1}) \big)  (\alpha_p^T\alpha_p)^{-1} \frac{\partial \alpha_p^T\alpha_p}{\partial \alpha_p}
\end{multline*}
with  
\begin{align*}
D(\alpha_p, p,x_{b,i},x_{d,i},b_i,d_i) = & \;\; 
-(d_{i}-b_{i})^2(\alpha_p^T \alpha_p) \\
+ \big( (x_{d,i}-p)(x_{d,i}-p)^T& d_i-(x_{b,i}-p)(x_{b,i}-p)^T b_i \big)(d_i-b_i)
\end{align*}




\begin{figure}
\centering
  \includegraphics[width=0.9 \linewidth]{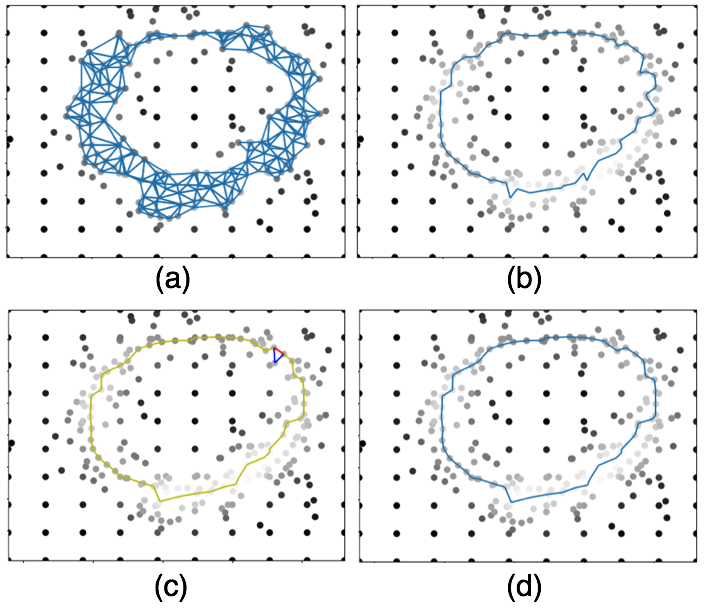}
  \caption{Steps of the Surface Reconstruction algorithm. (a) Simplicial complex (b) Generator/Proposed surface (c) Enumerating intersecting cofaces (d) 1D surface output}
  \label{fig:extract}
\end{figure}


This was implemented as a PyTorch module making experimentation with different cost functions straightforward.  As the backpropagation is separated for the diagrams,  the pipeline only requires an implementation of the derivative of the functional. Note that the cost function above is an illustrative example and the differentiable functions used to generate the examples are given in Section~\ref{sec:results}.

\section{Surface Reconstruction}
\label{sec:surface_recon}
The optimization in the previous section produces a likelihood function and a super-level set with the desired topology. In the ideal case, this super-level set is highly concentrated around the surface, but a surface with the correct topology must still be extracted. 
There are multiple ways to extract a 1D or 2D surface. One could apply the watershed algorithm~\cite{poranne20103d} to the inverse of the likelihood function, but then we lose control of the topology of the result. A further complication is that multiple super-level sets might need to be considered to flexibly locate the desired topology.

From a topological perspective, a natural candidate for the surface is a \emph{cycle representative} of the desired homology class. In practice we also consider the representative of the top dimension we consider (i..e 1-dimensional homology for curves and 2-dimensional homology for surfaces). 
As described in Section~\ref{sec:persistence}, we can associate a special representative, namely a Persistent Cycle Representative (PCR), with each point in the diagram. This is a $k$-dimensional cycle, which in our setting is either a 1-cycle (loop) or 2-cycle (sphere) and is part of the super-level set of the simplicial complex, $K^\alpha$. The following assumes the optimization was successful and that the PCR has the correct  topology.
For example in $\mathbb{R}^3$, we only consider 2-cycles such that the 2-cycles have the appropriate lower dimensional homology. If this is not the case, we declare that the optimization has failed to converge to a good local minimum.

The PCR, while topologically correct, is generally not the ideal geometric representation of the surface, e.g. it can be highly non-smooth and/or non-manifold. We therefore iteratively improve the PCR while ensuring that the correct topology is maintained. This is done through a local optimality function.  We use the number of simplices as a proxy for smoothness,  trying to  minimize the number of simplices in the cycle while maintaining topology. This is essentially improvement via \emph{simple homotopy}, a sequence of ``moves" which do not change the topology. 

\begin{figure}[tbp]
    \centering
    \includegraphics[width=\columnwidth]{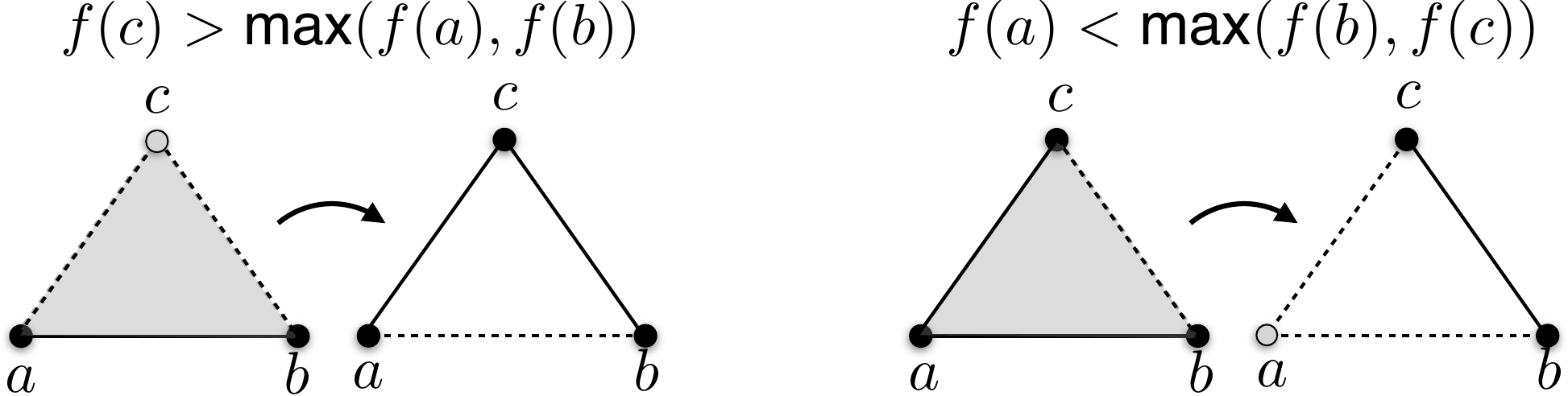}
    \caption{The two moves for modifying the candidate surface in 1D. (left) Only $\mathbf{a}$ and $\mathbf{b}$ are part of the surface. Vertex $\mathbf{c}$ has a higher function value than  $\mathbf{a}$ or $\mathbf{b}$ and so is added to the surface by replacing $(\mathbf{ab})$ by $(\mathbf{ac})$. (right) $\mathbf{a}$ has a lower function value than $\mathbf{b}$ or $\mathbf{c}$ and so is removed by replacing edges $(\mathbf{ac})$ and $(\mathbf{ab})$ by $(\mathbf{bc}$).}
    \label{fig:optimization1}
\end{figure}

Consider the 1D case, where we start with a 1-cycle PCR $g$. For each edge in $g$, we consider its adjacent triangles which lie in the superlevel set. There are two possible moves as shown Figure \ref{fig:optimization1}. Say we are considering $e=(a,b)$ and the triangle $(a,b,c)$. If neither $(b,c)$ nor $(a,c)$ are part of $g$, the triangle is added, replacing $(a,b)$ by the pair $(b,c)$ and $(a,c)$ if $c$ has a higher function value than both $a$ and $b$, i.e. $f(a)<\max(f(b),f(c))$. As the function is a proxy for a likelihood function, this has the effect of steering the path through a higher likelihood region. Alternatively, if $(a,c)$ is also part of $g$, the triangle addition essentially removes $a$ from the path, which only occurs  if $f(a)<\max(f(b),f(c))$, again pushing the path into a higher likelihood region.   
The procedure is illustrated in Figure~\ref{fig:extract}. At each step, we check that this does not cause any self-intersections and the algorithm terminates when no modification can be made. 
The algorithm works in any dimension and by construction the changes in the candidate surface do not change the topology. It also extends naturally to 2D surfaces, where we consider tetrahedra adjacent to triangles rather than triangles adjacent to edges. 

As mentioned, in Section~\ref{sec:persistence}, if there is no suitable representative (due to the requested topology), we output the simplicial complex associated with the appropriate super-level set with the correct topology.  The choice of the super-level set can be determined automatically from the persistence diagram. While it may occur that there is no super-level set with the appropriate topology, which again can be determined from the persistence diagram,  we say the algorithm has failed -- see Section ~\ref{sec:param}. In most cases, if the optimzation converged, a suitable super-level set could be found. We conclude that extracting a true surface in these cases is left for future work.


\section{Choice of Parameters and Failure Cases}\label{sec:param}

The parameters of the likelihood functions are initialized such that the derivative of the Gaussian is maximized at the average distance between points. This is done to make the basis functions as expressive as possible and empirically proved to be a good balance between a granular and a smoothed-out perspective. Other initializations that favored extreme and unexpected topologies did cause failure cases.
Furthermore, a weak point can be the fact that the map from the points in the persistent diagrams  to the complex, is not stable. In practice, this was not an issue. This could be explained by our desired topologies not being contingent on single simplices but instead on a larger set of simplices, making the ambiguity in the correspondences irrelevant. Lastly, large step sizes for gradient descent made the optimization process unstable, as expected, but it was straightforward to find a step size that worked for all our problems.

\section{Results}
\label{sec:results}

Our topology-aware surface reconstruction is novel in its flexibility in computing different topological reconstructions. 
Our technique also accepts topological information of any dimension. In this section, we discuss multiple results that validate our approach and compare it to other state-of-the-art surface reconstruction techniques. 

We remind the reader that in describing the cost functions, we use the convention that the points in the persistence diagram are sorted in decreasing lifetime, i.e. $|d_i-b_i| \geq |d_j-b_j| $ for $i<j$.



\subsection{Topological Flexibility}

Our technique can produce any topology that exists at any filtration value after the optimization process has finished. Thus, while it cannot always guarantee a certain topology, in practice and when the topology is not too far from those that are present in the original point cloud, our technique can be expected to produce it. This is true even if that exact topology is not present for any filtration value in the persistent homology of the original point cloud. 

\begin{figure}
\centering
  \includegraphics[width=\linewidth]{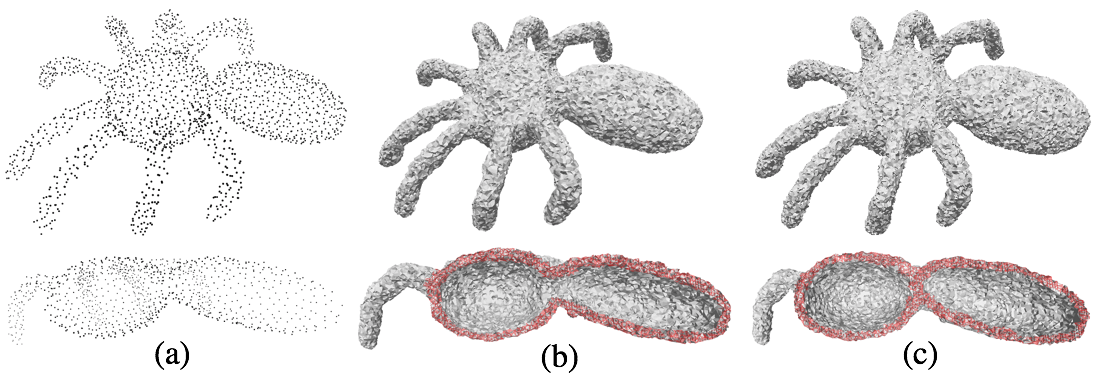}
  \caption{Reconstructing spider model with multiple possible topologies. (a) Input spider point cloud. (b) Creating 1 void inside the spider. (c) Creating 2 voids inside the spider. Cross sections in red.}
  \label{fig:top_guarantee_spider}
\end{figure}

To illustrate this flexibility, we show reconstructions from the same point cloud with different prescribed topologies.  In Figure~\ref{fig:top_guarantee_spider}, we see that a spider can be reconstructed to have one void or two. 
We used cost functions in $\mathrm{PD}(2)$, i.e. the  persistence diagram for 2-dimensional features. To reconstruct one void we used cost function:
$$
-((d_1-b_1)^2 - (d_2-b_2)^2)
$$
effectively maximizing the gap between the most persistent and the second most persistent homology classes. For two voids, we used the cost function: 
$$
-((d_2-b_2)^2 - (d_3-b_3)^2)
$$
maximizing the gap between the second and third most persistent classes. Most surface reconstruction techniques provide one major void, while the other is flattened out, while in our case, we can force this second void by means of predefining the requisite topology. 

\begin{figure}
\centering
  \includegraphics[width=\linewidth]{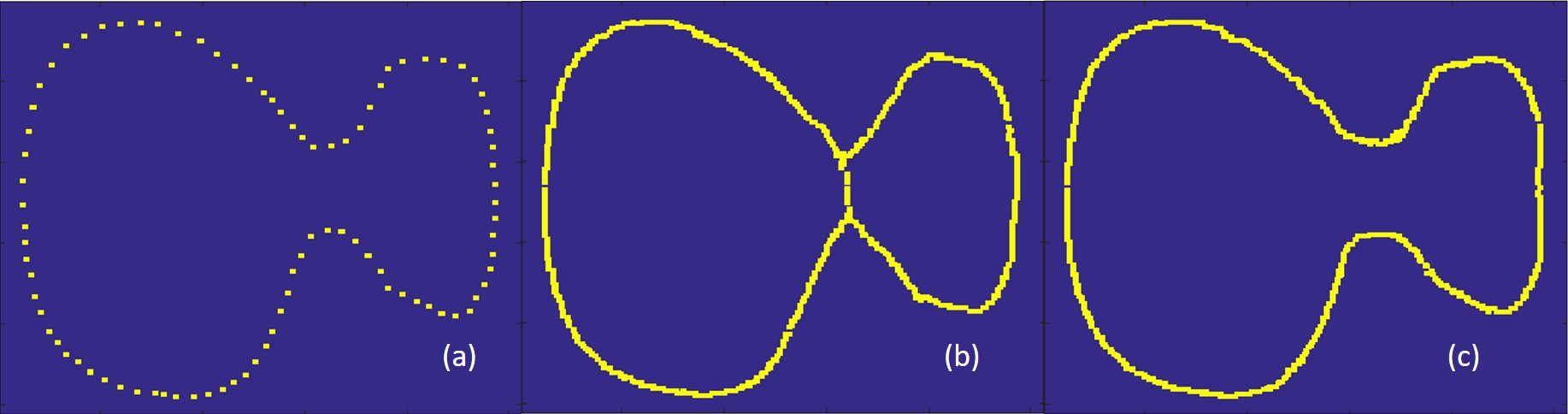}
  \caption{Reconstructing dual circle with multiple possible topologies. (a) The 2-D dual circle point cloud. (b) Reconstruction of dual circle as two adjacent circles. (c) Reconstruction of merged dual circle with one hole.}
  \label{fig:top_guarantee_ring}
\end{figure}

This topological flexibility can be further exemplified in the ring as shown in Figure~\ref{fig:top_guarantee_ring}. This is a two-dimensional case, with a point cloud as shown in Figure~\ref{fig:top_guarantee_ring}(a). The sampling makes it ambiguous whether the source shape contains one hole or two; but if this information is available, then our technique can be utilized to reconstruct the ring accordingly as in Figure~\ref{fig:top_guarantee_ring}(b) and (c) respectively. Most surface reconstruction techniques use tangent and normal information of points to reconstruct the underlying surface from them. This  means that the reconstruction with one hole would almost always be produced. Specifying the underlying topology beforehand can therefore be very useful.

\begin{figure}
\centering
  \includegraphics[width=\linewidth]{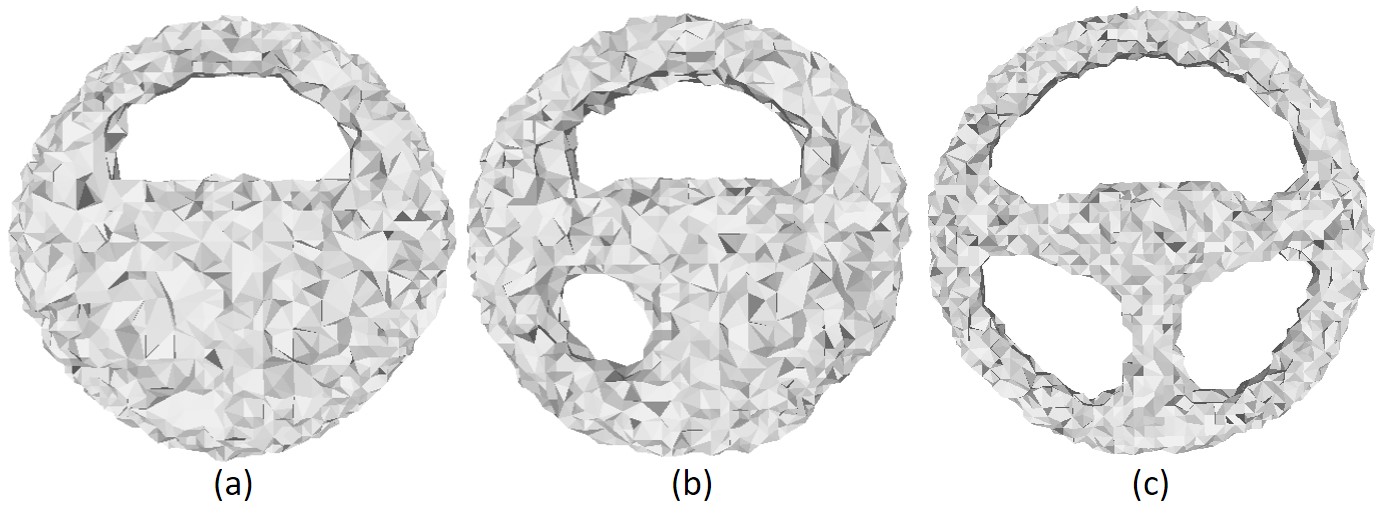}
  \caption{Reconstruction of steering wheel with multiple ring topologies. (a) 1 ring, (b) 2 rings, and (c) 3 rings}
  \label{fig:top_guarantee_steering}
\end{figure}

The steering wheel  in Figure~\ref{fig:top_guarantee_steering} provides another example. Here, different topologies (3,2 or 1 rings) are used to reconstruct an input point cloud, and the reconstructions are vastly different from each other. It is observed that for each of the different topologies, the technique progressively closes the less persistent ring. To achieve this, we used cost functions of the form
$$ 
-((d_k-b_k)^2 - (d_{k+1}-b_{k+1})^2)
$$
for $k=1,2,3$ taken over $\mathrm{PD}(1)$. By equipping the technique with a measure of topological preservation, it is able to obtain reconstructions that are seemingly very far off from the original point cloud or surface as in Figure~\ref{fig:top_guarantee_steering}(a) and (b), but respect topological requirements.


\subsection{Shape Completion}
\begin{figure}
\centering
  \includegraphics[width= \linewidth]{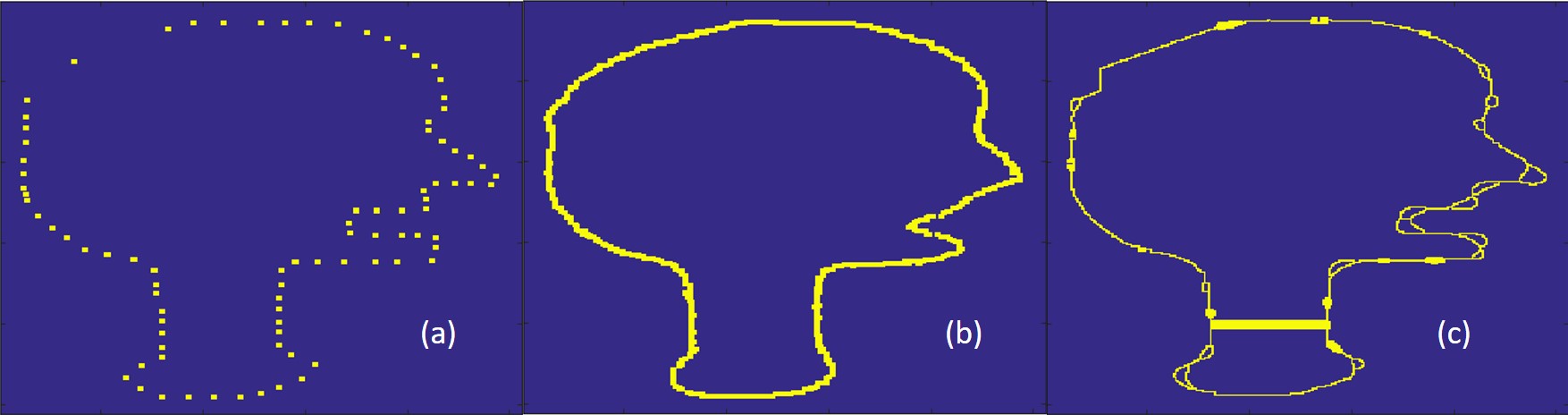}
  \caption{(a) 2-D incomplete head point cloud (b) Completing the point cloud with topological information ($1$-hole) (c) Using the generalized distance based watershed algorithm as in Poranne et al.~\cite{poranne20103d} to complete the point cloud.}
  \label{fig:completion_face}
\end{figure}

Shape completion is an important problem that we tackle from a topological perspective. 
We perform shape completion in both 2-dimensional and 3-dimensional examples. We show how an incomplete face  in Figure~\ref{fig:completion_face} is completed by means of our technique with the cost function: $-((d_1-b_1)^2)$ over $\mathrm{PD}(1)$. Here, the topology, i.e. one loop, of the face is provided as the only input apart from the input point cloud, and the technique performs the completion accordingly. It is also seen here that using the reconstruction technique described in Poranne et al.~\cite{poranne20103d} captures the high level structure of the shape but introduces a lot of topological noise. 

\begin{figure}
\centering
  \includegraphics[width=\linewidth]{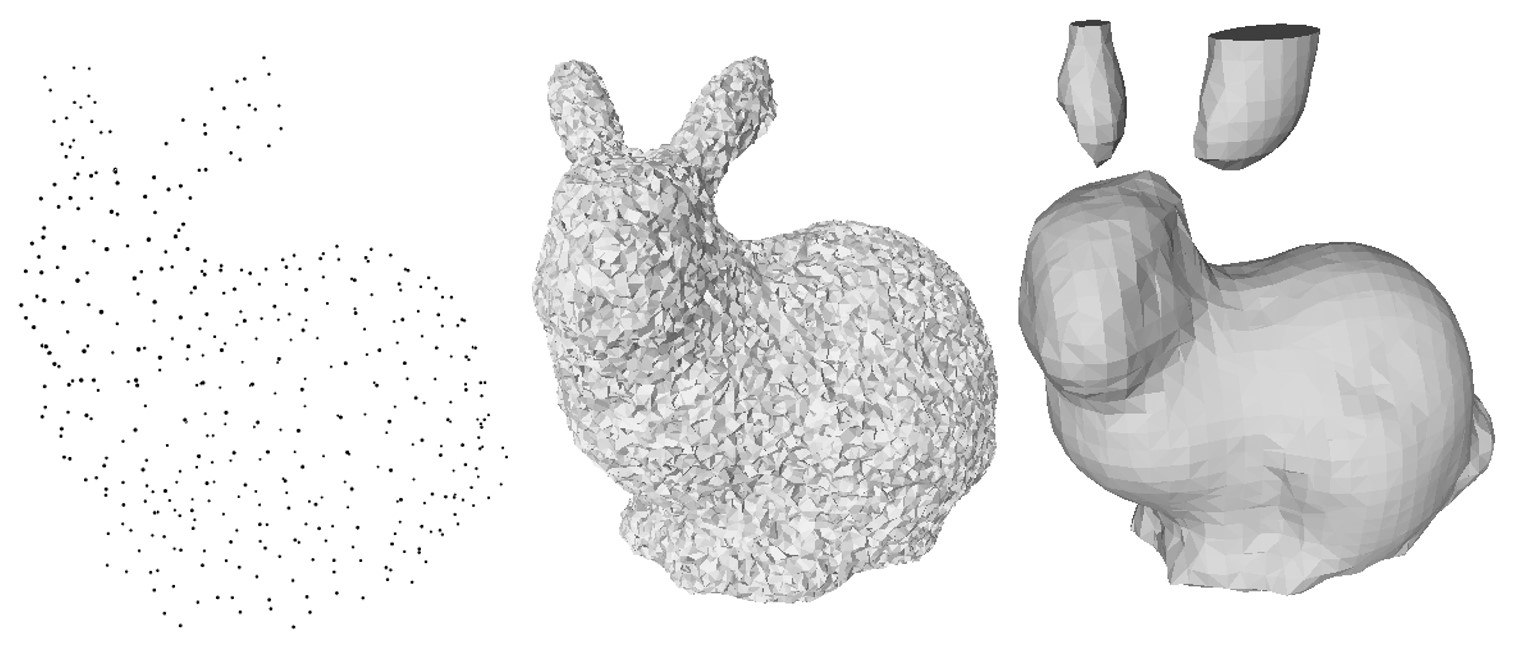}
  \caption{Densifying the Stanford bunny. (a) A sparse point cloud sampled from the Stanford bunny. (b) Reconstruction of this sampling, preserving only the essential 2-dim hole (c) Poisson reconstruction of the sparse Stanford bunny.}
  \label{fig:noise_bunny}
\end{figure}

Another situation where our technique proves useful is when sparse sampling introduces topological ambiguity. In Figure~\ref{fig:noise_bunny}(a), there are too few points in the input point cloud. By reconstructing the surface with a single 2-dimensional hole, it forces the reconstruction to produce a consistent, bunny-like reconstruction, Figure~\ref{fig:noise_bunny}(b), using the cost function: 
$ -((d_1-b_1)^2-(d_2-b_2)^2)$ taken over $\mathrm{PD}(2)$. This is not necessarily the case when other reconstruction techniques are used. For example, the Poisson surface reconstruction of the sparse Stanford bunny can be seen in Figure~\ref{fig:noise_bunny}(c). The sparsity of the point cloud creates three major concentrations of points in the shape, therefore producing a topologically incorrect reconstruction.

\subsection{Medical Data}

One of the central motivations for topology-aware surfaces is the potential application to medical data. As described in Section~\ref{sec:intro}, the topology of medical data is almost always predefined and invariant across numerous instances. This is a property that can be utilized, along with other medical priors, to reconstruct various medical organs from scan data. 

\begin{figure}
\centering
  \includegraphics[width=\linewidth]{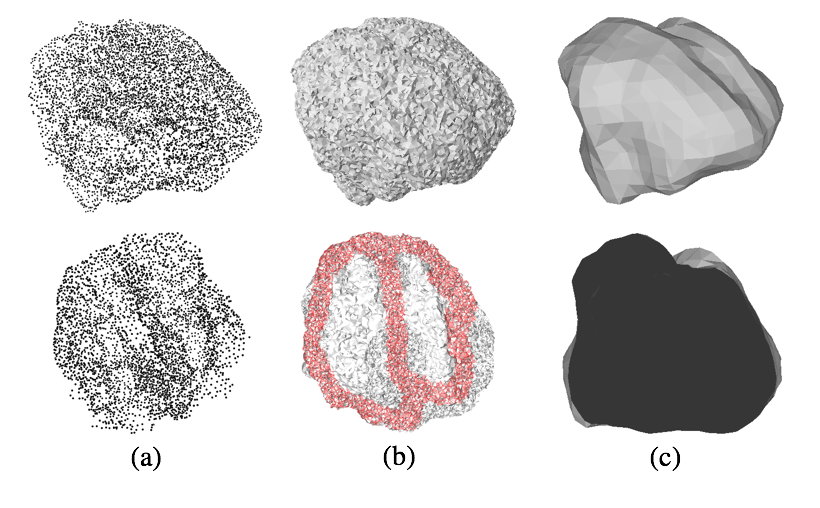}
  \caption{Reconstruction of the human brain as seen from the top-view. (a) A point cloud of an example brain from Lerma-Usabiaga et al.~\cite{lerma2018converging}. (b) Reconstruction of this brain, preserving two central voids. (c) Poisson Reconstruction of the brain. Cross sections in red.}
  \label{fig:brain}
\end{figure}

Here we show examples of how one could use our technique to reconstruct scans of different medical organs. In Figure~\ref{fig:brain}, we show the reconstruction of the brain from the scan data provided in Lerma-Usabiaga et al.~\cite{lerma2018converging}. We used the cost function: $$ -((d_2-b_2)^2-(d_3-b_3)^2) $$ over $\mathrm{PD}(2)$. In this case, even though the input point cloud is dense, it is fairly complex to reconstruct. While our technique reconstructs the brain fairly close to the input point cloud, the more classical Poisson surface reconstruction~\cite{kazhdan1poisson} fails to obtain a topologically correct reconstruction, producing a single three-dimensional void, while the correct topology would consist of two three-dimensional voids, one for each hemisphere of the brain.

\begin{figure}
\centering
  \includegraphics[width=\linewidth]{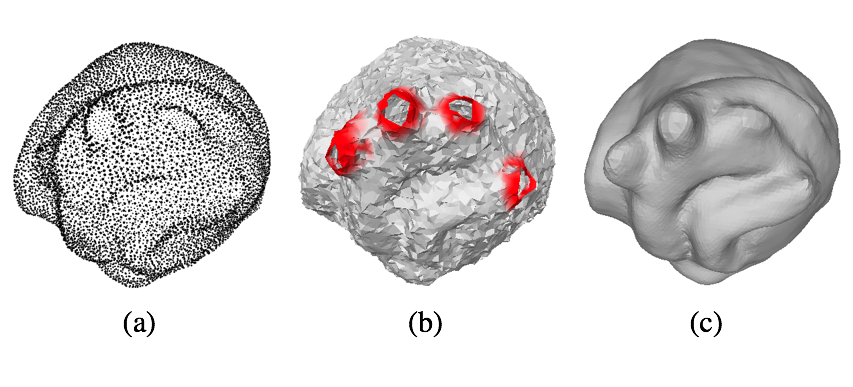}
  \caption{Reconstruction of the human heart model. (a) A point cloud of a human heart model. (b) Our reconstruction with four openings; the red markings here indicate the four holes in the reconstruction. (c) Poisson reconstruction.}
  \label{fig:heart}
\end{figure}

We also attempted to reconstruct synthetic point clouds of human hearts.
The reconstruction in Figure~\ref{fig:heart} was obtained using the cost function: $$ -((d_3-b_3)^2-(d_4-b_4)^2) $$ over $\mathrm{PD}(1)$. While our method reconstructs the four openings in the human heart, Poisson reconstruction smooths the holes, changing the topology of the reconstructed surface.
\subsection{Comparisons to other Surface Reconstruction techniques}
 We compare our topology-aware technique to two major surface reconstruction techniques. For 2D surface reconstruction, we compare our technique to that of Poranne et al.~\cite{poranne20103d}, and for 3D surface reconstruction, we compare primarily to Poisson surface reconstruction~\cite{kazhdan1poisson} as implemented in MeshLab. These qualitative comparisons can be seen in Figures~\ref{fig:comparison_poisson} and~\ref{fig:completion_face}. While the Poisson reconstruction is very good on smooth water-tight surfaces, it tends to fail for more natural cases where the inherent topology is non-manifold. This is observed in the non-manifold examples of Figure~\ref{fig:comparison_poisson} such as chair, cup, octopus, and table examples, where the Poisson reconstruction attempts to obtain a water-tight surface and fails both topologically and in reconstruction quality. We perform our 2D quantitative comparison experiments on the 2D point clouds used in Poranne et al.~\cite{poranne20103d}, specifically the \textit{hand}, \textit{spiral}, \textit{helix}, \textit{face}, \textit{circle} and \textit{blob} datasets, and the 3D comparisons on 8 categories of the McGill segmentation benchmark dataset~\cite{siddiqi2008retrieving}.


We develop two different metrics to compare with existing algorithms. We compare against Poisson surface reconstruction~\cite{kazhdan1poisson} for 3D reconstruction and Generalized distance based reconstruction~\cite{poranne20103d} for both 2D and 3D reconstruction. Since our technique aims at topological accuracy, we compare against the topological accuracy of other techniques. This is done by computing the average number of erroneous $k$-dimensional components: $\text{TFI}_k = \frac{1}{N}\sum_{i=1}^N |n_{i,k}-n_{i,k}^\text{recon}|$, where $N$ refers to the number of shapes. We refer to $\text{TFI}_k$ as the $k-$dimensional Topological Fidelity Index. Here, if $k=1$, then $n_{i,1}$ is the number of true $1$-D holes in shape $i$, $n_{i,k}^\text{recons}$ refers to the number of $1$-D holes in the reconstruction of shape $i$. We present this for multiple scan sizes of our input point cloud. This is presented in Table~\ref{tab:top}. Our topological flexibility ensures a TFI of $0$ for our technique, while a bigger TFI value for other techniques is a measure of how far from ideal their reconstruction capabilities are topologically. The high TFI values for both competing methods shows the value of our method that provides topological flexibility.

\begin{table}
\begin{center}
\begin{tabular}{|p{3cm}|c|c|c|}
\hline
\textbf{Method - 2D} & $N=1000$ & $N=500$ & $N=200$ \\
\hline\hline
Ours (k=0) & \textbf{0} & \textbf{0} & \textbf{0} \\
Watershed~\cite{poranne20103d} & 0.25 & 0.33 & 0.81 \\
\hline
Ours (k=1) & \textbf{0} & \textbf{0} & \textbf{0} \\
Watershed & 2.25 & 3.08 & 3.63 \\
\hline\hline
\textbf{Method - 3D} & $N=1000$ & $N=500$ & $N=200$ \\
\hline\hline
Ours (k=0) & \textbf{0} & \textbf{0} & \textbf{0} \\
Watershed~\cite{poranne20103d} & 0.25 & 0.375 & 0.375 \\
Poisson~\cite{kazhdan1poisson} & 0.75 & 1.125 & 0.25 \\
\hline
Ours (k=1) & \textbf{0} & \textbf{0}& \textbf{0} \\
Watershed & 1.625 & 2 & 2.125 \\
Poisson & 0.875 & 0.875 & 0.875 \\
\hline
Ours (k=2) & \textbf{0} & \textbf{0} & \textbf{0} \\
Watershed & 0.375 & 0.375 & 0.375 \\
Poisson & 0.5 & 0.5 & 0.625 \\
\hline
\end{tabular}
\end{center}
\caption{$\text{TFI}_0,\text{TFI}_1$ (and $\text{TFI}_2$) for $2$D (and $3$D) surface reconstruction comparing our technique to the generalized-distance Watershed algorithm by Poranne et al.\cite{poranne20103d} (and the Poisson surface reconstruction by Kazhdan et al.\cite{kazhdan1poisson} respectively) for point clouds of size $N$ points. The lowest values are captured in bold.}
\label{tab:top}
\end{table}
\begin{table}
\begin{center}
\begin{tabular}{|p{3cm}|c|c|c|}
\hline
\textbf{Method - 2D} & $N=1000$ & $N=500$ & $N=200$ \\
\hline\hline
Ours & \textbf{1.0040} & \textbf{1.0026} & \textbf{0.9979} \\
Watershed~\cite{poranne20103d} & 1.0363 & 1.0479 & 1.0723 \\
\hline\hline
\textbf{Method - 3D} & $N=1000$ & $N=500$ & $N=200$ \\
\hline\hline
Ours & \textbf{1.9170} & \textbf{1.9143} & \textbf{1.9127} \\
Watershed~\cite{poranne20103d} & 1.9867 & 1.9828 & 1.9800 \\
Poisson~\cite{kazhdan1poisson} & 2.0348 & 2.0309 & 2.0117 \\
\hline
\end{tabular}
\end{center}
\caption{One-way Chamfer distance for $2$D (and $3$D) surface reconstruction comparing our technique to the generalized-distance Watershed algorithm~\cite{poranne20103d} (and the Poisson surface reconstruction~\cite{kazhdan1poisson} respectively) for point clouds of size $N$ points. The lowest values are captured in bold.}
\label{tab:chamfer}
\end{table}


While the TFI captures topological fidelity, another important factor is the quality of reconstruction. To capture this, we compute the one-way Chamfer distance between a point cloud and its reconstruction. This score is normalized for all shapes in a collection and averaged over the collection. The comparison of these scores across the corresponding techniques for 2D and 3D is provided in Table~\ref{tab:chamfer}. Here we show that our technique, in addition to high topological fidelity, also obtains high quality surface reconstruction.

%
%


\section{Conclusion and Future Work}
\label{sec:conclusion}

In this work we introduce a technique that performs surface reconstruction from point scans while optimizing the topology of the surface. The results show that the optimization converges to the desired topology and yields accurate reconstructions.  There are many potential future directions for improvement. 
For example, while our choice of basis and cost functions are intuitive and work well in practice, there are many other possibilities which could be explored. While we do not have guarantees on the reconstruction, if the functional does not converge or converges to a large value, we can detect the failure. Likewise, if the extracted generator does not have the correct topology or is far from the points, we can again detect failure. As our algorithm is based on gradient descent we can retry with different initializations.  

Our approach for extracting the surface from the super-level set and subsequently optimizing it is currently quite simple and could clearly be improved. As the examples show, the result is ``close" and has the correct topology, the results would be improved if local geometric features such as smoothness could be taken into account. This could produce much higher resolution reconstructions while remaining ``close" to the topologically correct surface. This could also take normal information into account which could further improve both the optimization and the final extracted surface. An alternative approach would be to use the super-level set as a guide or  constraint on implicit methods to extract the final surface. It remains an open question how  implicit methods could be modified to extract non-manifold surfaces, e.g. manifolds with boundaries, intersections of manifolds, etc. This is a future direction which we intend to pursue.

The general area of topology-aware geometry processing remains largely unexplored, with many possible future directions of research. Here we addressed only single scan reconstructions while it may be possible to jointly optimize over a collection as shape categories in collections often  share many topological properties. Similarly, this type of approach could also be applied to time-varying scans, e.g. point clouds representing motion, where the codimension of the surface is greater than 1. Finally, in the case of collections our approach could be extended to combine  topology inference with reconstruction or representation learning by inferring the topology at the same time as optimizing with respect to it. 

\section{Acknowledgements}

Supported by the Slovenian Research Agency (ARRS N1-0058) and by the EU H2020-MSCA-RISE project RENOIR (grant no. 691152). Supported by Altor Equity Partners AB through Unbox AI (unboxai.org). We would like to acknowledge the help of Garikoitz Lerma Usabiaga for his help in understanding brain scan data and manipulating it through the Freesurfer software.

\printbibliography                


\end{document}